\documentclass[preprint,12pt]{elsarticle}

\usepackage{amssymb}
\usepackage{multirow}
\usepackage{booktabs}
\usepackage{amssymb}
\usepackage{mathrsfs}
\usepackage{longtable}
\usepackage{lscape}
\usepackage{tabularx}
\usepackage{epsfig}
\usepackage{colortbl}
\journal{arXiv.org}

\begin{document}

\begin{frontmatter}



\title{A short note on the axiomatic requirements of uncertainty measure}


\author[swu]{Xinyang Deng}
\author[swu,vu]{Yong Deng\corref{cor}}
\ead{ydeng@swu.edu.cn, prof.deng@hotmail.com}

\cortext[cor]{Corresponding author: Yong Deng, School of Computer and Information Science, Southwest University, Chongqing, 400715, China.}

\address[swu]{School of Computer and Information Science, Southwest University, Chongqing, 400715, China}
\address[vu]{School of Engineering, Vanderbilt University, Nashville, TN, 37235, USA}

\begin{abstract}
In this note, we argue that the axiomatic requirement of \emph{range} to the measure of aggregated total uncertainty (ATU) in Dempster-Shafer theory is not reasonable.
\end{abstract}

\end{frontmatter}

Dempster-Shafer theory \cite{Dempster1967,Shafer1976} is widely applied to uncertainty modeling \cite{Denoeux2010479,Smets1994191}. Two types of uncertainty, namely nonspecificity and discord, are coexisting in the Dempster-Shafer theory \cite{Klir1995,jousselme2006measuring}. A justifiable measure to these uncertainty is necessary to describe the essential characters of basic probability assignment function (BPA). To be justifiable, for a measure called as aggregated total uncertainty (ATU), some requirements are necessary. In the previous study, an ATU is required to satisfy five axiomatic requirements \cite{klir1999uncertainty,klir2008remarks}. One of the five axiomatic requirements is the requirement of \emph{range}, which is shown as follows \cite{klir2008remarks}.

\begin{itemize}
\item \emph{Requirement of range: When ATU is measured in bits, its range must be consistent with the conceivable limits of uncertainty to be found in a set that is the size of $X$, which means that}
\begin{equation}\label{Eq1}
0 \le ATU(m) \le \log _2 |X|
\end{equation}
\emph{where $m$ denotes a BPA, and $X$ denotes a frame of discernment.}
\end{itemize}

According to the requirement, a BPA's uncertainty is limited in the range of $[0, \log _2 |X|]$. However, it is not reasonable. Some examples are given as below.

Suppose there are 32 students who participated in a course examination. After the examination finished, a student won the first place. In order to know who is the first one, we has to ask their course teacher. But the teacher doesn't want to directly tell us who is the first one. Instead, she just answers ``Yes" or ``No" to our question. The problem is how many times do we have to ask at most in order to know who \textbf{IS} the first \textbf{ONE}?

Assume the times is $t$. It is easy to answer the problem by the definition of Shannon's entropy
\begin{equation}
t = \log _2 {32} = 5
\end{equation}

\begin{figure}[!t]
\centering
\includegraphics[angle=-90,width=6.5in]{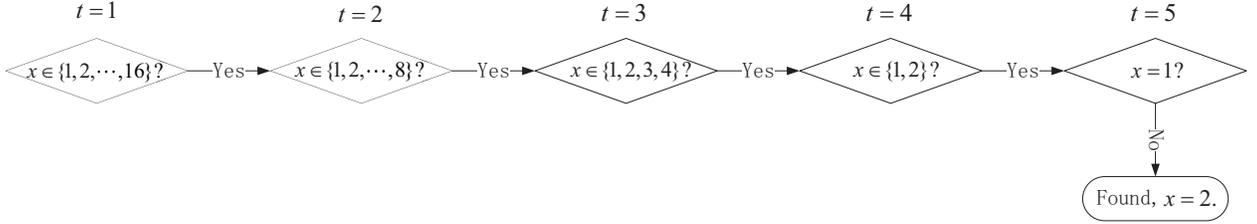}
\caption{Process to find the top 1 student}
\label{fig1}
\end{figure}

For example, assume the first student's number is No. 2. To find this student, we suppose the first student's number is $x$. According to the process shown in Fig.\ref{fig1}, we can find the top 1 student is No.2 through five times asking at most.

Now, let's consider another situation. Assume we have been told that there are two students tied for first. In this case, how many times do we have to ask at most to know who \textbf{ARE} the first \textbf{ONES}?

In this case, obviously
\begin{equation}
t \geq \log _2 {32}
\end{equation}

What's more, in the most extreme case assume there are 32 students tied for first. However, we don't have any information (\textbf{TOTALLY UNKNOWN}) about this examination. In order to know who is (are) the first one (ones), we have to ask $\log _2 2^{32} = 32$ times.

Let's consider another case. Given a frame of discernment $X = \{A, B, C \}$, if we don't have any information, it can be expressed as $m(A, B, C) = 1$. If a piece of information is obtained that $A$, $B$ and $C$ are equally happened, it can be expressed as $m(A) = m(B) = m(C) = 1/3$ (In this situation, the BPA is degenerated to a probability distribution). Intuitively, the uncertainty degree in the situation of TOTALLY UNKNOWN that $m(A, B, C) = 1$ is larger than that of the situation of EQUALLY HAPPENED that $m(A) = m(B) = m(C) = 1/3$.

As discussed above, we draw a conclusion that the upper bound shown in Eq. (\ref{Eq1}) is not reasonable in the Dempster-Shafer theory.

%



\bibliographystyle{elsarticle-num}
\bibliography{References}







\end{document}